\documentclass[twocolumn]{aastex62}
\usepackage{amsmath}
\usepackage{bm}
\usepackage{color}

  \received{Jan. 18, 2018}
  \revised{Apr. 23, 2018}
  \accepted{Apr. 27, 2018}
  \submitjournal{AJ}

  \shorttitle{Retrograde 1:1 mean motion resonance}
  \shortauthors{Yukun Huang et al.}

\begin{document}

\title{Dynamic portrait of the retrograde 1:1 mean motion resonance}

\author{Yukun Huang}
\affil{School of Aerospace Engineering, Tsinghua University \\
	Beijing, China, 100086}

\author{Miao Li}
\affiliation{School of Aerospace Engineering, Tsinghua University \\
	Beijing, China, 100086}

\author{Junfeng Li}
\affiliation{School of Aerospace Engineering, Tsinghua University \\
	Beijing, China, 100086}

\author{Shengping Gong}
\affiliation{School of Aerospace Engineering, Tsinghua University \\
	Beijing, China, 100086}



\begin{abstract}
	Asteroids in mean motion resonances with giant planets are common in the solar system, but it was not until recently that several asteroids in retrograde mean motion resonances with Jupiter and Saturn were discovered. A retrograde co-orbital asteroid of Jupiter, 2015 BZ509 is confirmed to be in a long-term stable retrograde 1:1 mean motion resonance with Jupiter, which gives rise to our interests in its unique resonant dynamics. In this paper, we investigate the phase-space structure of the retrograde 1:1 resonance in detail within the framework of the circular restricted three-body problem. We construct a simple integrable approximation for the planar retrograde resonance using canonical contact transformation and numerically employ the averaging procedure in closed form. The phase portrait of the retrograde 1:1 resonance is depicted with the level curves of the averaged Hamiltonian. We thoroughly analyze all possible librations in the co-orbital region and uncover a new apocentric libration for the retrograde 1:1 resonance inside the planet's orbit. We also observe the significant jumps in orbital elements for outer and inner apocentric librations, which are caused by close encounters with the perturber.


\end{abstract}



\section{Introduction}\label{sec:1}

The recently discovered asteroid 2015 BZ509, the first asteroid in retrograde 1:1 mean motion resonance (or retrograde 1:1 resonance) with Jupiter, has arguably been the most intriguing small body inside the Solar System \citep{Morais:2017fn}. It was first detected in January 2015 and was confirmed to be in the retrograde 1:1 resonance with Jupiter by \citet{Wiegert:2017fj} in 2017. It is shown that 2015 BZ509 has long-term stability for around a million years in numerical integration. However, the origin of 2015 BZ509 is still not clear. It could be a Democloid or Halley family comet that was captured into retrograde resonance, but no substantial evidence can prove that guess.

Except for the mysterious 2015 BZ509, other asteroids in retrograde resonance with Jupiter or Saturn have also been found. \citet{Morais:2013ka} confirms that two asteroids, 2006 BZ8 and 2008 SO218 are currently in retrograde resonance with Jupiter, while another one, 2009 QY6 is in retrograde resonance with Saturn. Additionally, there are two asteroids, 2006 BZ8 and 1999 LE31, likely to be captured into retrograde resonance with Saturn.

Not only in our solar system, \citet{2008A&A...482..665G} suggested that retrograde resonances are possible in multi-planet systems
using MEGNO method proposed by \citet{Cincotta:2000ua}. It was shown that the two planets of the exoplanetary system HD 73526 are possibly in a retrograde 2:1 mean motion resonance. \citet{2009Icar..201..381S,2010CeMDA.107..487S} demonstrated numerically that retrograde planets can survive in more closely packed configurations than prograde planets. After that, the possibility of planetary retrograde configuration is revalidated by comparing the stability of retrograde and prograde planet in a circular binary system \citep{2012MNRAS.424...52M}. It seems that a retrograde system is surprisingly more stable than its prograde counterpart.

The motion of asteroids in mean motion resonance with a planet, especially with Jupiter, has been studied for about two centuries \citep{Murray:2000id, Morbidelli:2002tn}. The analytical tools of mean motion resonance can handle most of the case, while, retrograde one excluded. This is because retrograde asteroids are far rarer than prograde asteroids, not much attention was paid to these odd small bodies for decades. Until now, only 93 discovered asteroids are confirmed to be retrograde\footnote{JPL Small-Body Database Search Engine (\url{https://ssd.jpl.nasa.gov/sbdb_query.cgi}), retrieved on January 2018.}.

Date back to the last century, probably the earliest mention of retrograde mean motion resonance was undertaken by \citet{Schubart:1978ass}, in which certain initial values corresponding to the retrograde 1:1 resonance with Jupiter were numerically investigated. It was shown that the motion of the test particles is stable and they will not come too close to Jupiter. So far, numerous numerical experiments have provided us good maps to understand the dynamical structure of the retrograde resonance. The stability map of coorbital resonance depicts the resonant and chaotic areas of retrograde 1:1 resonance in both planar and 3D case \citep{Morais:2016kq}. It is shown that capture mechanisms for 2D and 3D retrograde orbit are intrinsically different. The probabilities of capture into resonance at different inclinations are studied numerically by \citet{Namouni:2014dz,Namouni:2017dh}. Surprisingly, the retrograde resonances are more efficient at capture than prograde resonances. The absolute capture probability of the coorbital capture was investigated with statistical simulations in \citet{Namouni2017}. Particularly, it is worth noting that 2015 BZ509 sits near the peak of coorbital capture efficiency, while the reason for that is still unclear without a clear sight into the phase-space structure of the retrograde 1:1 resonance.

To get a clear and comprehensive understanding of the dynamical structure of retrograde resonance, numerical experiments alone are not adequate. Analytical tools are necessary to handle this problem. \citet{Gayon:2009ct} applied an analytical method based on Hamiltonian approach and expansion of the disturbing function to a general three-body system. The theory is used to address their interested planetary retrograde resonance problem and shown to be a good match for the numerical integration. \citet{2013CeMDA.117..405M} demonstrated that a retrograde resonance problem can be transformed to another equivalent prograde problem with variable substitution. Then the classical method addressing the prograde problem \citep[Chapter 8]{Murray:2000id} can be applied to the retrograde case.

Whereas for the retrograde 1:1 resonance, the series expansion of the disturbing function is no longer valid since its ratio of semi-major axes is near unity \citep{2013CeMDA.117..405M}. While considering other mean motion commensurabilities like 2:1 or 3:1, the series expansion is only valid for small eccentricity. When the eccentricity of the asteroid is above the planet-crossing limit ($e>1 / \alpha-1$ for outer perturber and $e>1 - 1 / \alpha$ for inner perturber), a new libration zone arises \citep{Moons:1993fv,2017AJ....154...20W}, which is now explained as the phase protection mechanism from planetary collisions provided by mean motion resonance \citep{Morbidelli:2002tn}. To sum up, the series expansion can only be used when orbits under consideration do not cross \citep{2017AJ....154..229B,Mardling:2013ba} and the ratio of semi-major axes is never unity.


In light of the deficiency of series expansion approach, \citet{2013CeMDA.117..405M} proposed a semi-analytic model based on ponderomotive potential to study the retrograde 1:1 resonance. They show that pericentric libration (libration around $0^\circ$) and apocentric libration (libration around $180^\circ$) both exist, and the collision separatrix play a big role in the retrograde 1:1 resonance. It is the first time that the dynamical structure of the planar retrograde 1:1 resonance has been unveiled. However, the study is limited to the case where semi-major axis $a = 1.01$ and does not provide a global and comprehensive phase-space portrait of the retrograde 1:1 resonance.


In this paper, we follow another procedure as described in \citet{Morbidelli:2002tn,Moons:1993fv} and adapt it to the retrograde case. In Section~\ref{sec:2}, we employ a series of canonical contact transformation \citep[Chapter 2]{Murray:2000id} to obtain a simple integrable approximation for the planar retrograde resonance. In Section~\ref{sec:3}, we concentrate on the most complicated retrograde resonance, the retrograde 1:1 resonance and carry out the averaging procedure numerically in closed form. The level curves of the averaged Hamiltonian (i.e., phase portraits of the resonance) are depicted. We globally analyze the phase-space structure of the retrograde 1:1 resonance in detail, especially taking into account the effect of the collision curve.


\section{Retrograde MMR integrable approximation}\label{sec:2}
This section is devoted to the construction of a general analytical theory for all the retrograde mean motion resonances. Firstly, we introduce the following retrograde Poincare variables \citep{Gayon:2009ct, Namouni:2014dz}
\begin{equation}\label{eq:retrograde_poincare_variables}
	\begin{aligned}
		\Lambda^* & = L,\quad     & \lambda^* & = M + \omega - \Omega,         \\
		P^*       & = L - G,\quad & p^*       & = -\left(\omega-\Omega\right), \\
		Q^*       & = G + H,\quad & q^*       & = \Omega,                      \\
	\end{aligned}
\end{equation}
where $L = \sqrt{a},\ G = L\sqrt{1-e^2},\ H = G\cos{i}$ and $M,\omega,\Omega$ are mean anomaly, argument of perihelion and longitude of node respectively. The canonical variables $L, M, G, \omega, H, \Omega$ are usually called the Delaunay variables.
It is easy to check that the following contact transformation relationship holds:
\begin{equation}\label{eq:contact_transformation}
	\Lambda^*\lambda^* + P^*p^* + Q^*q^* = LM+G\omega+H\Omega.
\end{equation}
So the new retrograde Poincare variables is a set of canonical action-angle variables.

The reason to modify the classic Poincare variables for the retrograde case is simple. When we focus on prograde orbit, the third action of Poincare variables is $Q=G-H=2G\sin^2{\frac{i}{2}}$ and $Q$ approaches zero as $i$ approaches $0^\circ$. This property of $Q$ makes the study of planar mean motion resonance easy. Therefore we want the similar property for our new variables. The most intuitive way is to change the definition of $Q$ to $Q^* = G + H = 2G\cos^2{\frac{i}{2}}$. And at the same time, the definition of angle variables must also be changed in order to satisfy the contact transformation relationship.

The major difference between the angle variables is that the definition of $\varpi^*$ is now $\omega-\Omega$ rather than $\omega+\Omega$. Though $\varpi^*$ is not the longitude of perihelion anymore (to be precise, it is now the negative longitude), we still use this terminology for simplicity.
With the new definition of $\varpi^*$ in mind, we have the same relation between $\lambda^*$ and $\varpi^*$ that $\lambda^* = M + \varpi^*$, just like prograde orbit. And again, $\lambda^*$ is the negative mean longitude of a retrograde orbit.

From now on, we remove the symbol $^*$ that denotes retrograde orbits for the sake of simplicity, and it is important for readers to keep in mind that $\varpi = \omega - \Omega$ and $\lambda = M + \omega - \Omega$ for the retrograde asteroid.

A retrograde mean motion resonance between an asteroid and a planet (denoted by superscript symbol $^\prime$) occurs when $kn-k^\prime n^\prime \approx 0$, where $k$ and $k^\prime$ are both positive integers. We call such resonance a retrograde $k^\prime : k$ resonance. $n$ and $n^\prime$ are the mean motion frequencies of the retrograde asteroid and the planet respectively and both mean motions are considered positive. The mean motion resonant normal form Hamiltonian for the circular restricted three-body model is given by \citep[Chapter 9]{Morbidelli:2002tn}
\begin{equation}\label{eq:mean_motion_resonant_normal_form_hamiltonian}
	\mathcal{H}_{MMR} = \mathcal{H}_0\left(\Lambda,\Lambda^{\prime}\right) + \epsilon \mathcal{H}_1\left(\Lambda, P, Q, p, q, k \lambda- k^\prime \lambda^\prime\right),
\end{equation}
where
\begin{equation}
	\mathcal{H}_0 = -\cfrac{1}{2\Lambda^2} + n^\prime \Lambda^\prime
\end{equation}\label{eq:normal_form_hamiltonian_main_term}
is the main term, and $\mathcal{H}_1$ is the disturbing term. It is easy to see from D'Alembert rules, that the disturbing function depends only on two angles. So we introduce another set of canonical action-angle variables:
\begin{equation}\label{eq:retrograde_canonical_variables}
	\begin{aligned}
		S                        & = P,                                             & \sigma                   & = \cfrac{k \lambda- k^\prime \lambda^{\prime}+\left(k+k^\prime\right)p}{\left(k+k^\prime\right)}, \\
		S_z                      & = Q,                                             & \sigma_z                 & = \cfrac{k \lambda- k^\prime \lambda^{\prime}+\left(k+k^\prime\right)q}{\left(k+k^\prime\right)}, \\
		N                        & =-\cfrac{k+k^\prime}{k} \Lambda + P +Q,          & \nu                      & = -\cfrac{k \lambda- k^\prime \lambda^{\prime}}{\left(k+k^\prime\right)},                         \\
		\tilde{\Lambda}^{\prime} & = \Lambda^{\prime} + \cfrac{k^\prime}{k}\Lambda, & \tilde{\lambda}^{\prime} & = \lambda^{\prime},
	\end{aligned}
\end{equation}
where $\sigma$ and $\sigma_z$ are the critical angles of the retrograde mean motion resonance and they were also found by \citet{2013CeMDA.117..405M}. And the Hamiltonian in the new variables turn out to be:
\begin{equation}\label{eq:new_hamiltonian_in_resonant_variables}
	\begin{aligned}
		\mathcal{H}_{MMR} = & \ \mathcal{H}_0\left(\tilde{\Lambda}^{\prime},N,S,S_z\right)                                                \\
		                    & + \epsilon \mathcal{H}_1\left(N,S,S_z,\left(k^\prime+k\right)\sigma,\left(k^\prime+k\right)\sigma_z\right).
	\end{aligned}
\end{equation}
It is a system with two degrees of freedom that is in general nonintegrable. To get an integrable approximation of the Hamiltonian, first we restrict our retrograde motion to the planar case $i=180^\circ$ and then $S_z$ or $Q$ is always equal to $0$. As a result, the corresponding angle variables $\sigma_z$ can be eliminated from equation~\eqref{eq:new_hamiltonian_in_resonant_variables} \citep[Equation (9.4)]{Morbidelli:2002tn}. Besides, it is evident that $\tilde{\Lambda}^{\prime}$, $N$ are constants of motion for $\mathcal{H}_{MMR}$. Thus we can drop the term $n^\prime\tilde{\Lambda}^{\prime}$ from $\mathcal{H}_{MMR}$ and the Hamiltonian of the planar case now only depends on one critical angle $\sigma$. The new planar Hamiltonian reads:
\begin{equation}\label{eq:planar_hamiltonian}
	\mathcal{H}_{PC} = \mathcal{H}_0\left(N,S\right) + \epsilon \mathcal{H}_1\left(N,S,\left(k^\prime+k\right)\sigma\right),
\end{equation}
where
\begin{equation}\label{eq:planar_hamiltonian_main_term}
	\mathcal{H}_0 = -\cfrac{{\left(k^\prime+k\right)}^2}{2k^2{\left(N-S\right)}^2} + n^\prime \cfrac{k^\prime}{\left(k^\prime+k\right)}\left(N-S\right).
\end{equation}
With a planar integrable system in hand, we can then plot the level curves of the $\mathcal{H}_{PC}$ on the $S$, $\sigma$ plane, for different values of N. As for the disturbing term $\epsilon \mathcal{H}_1$, we numerically evaluate its value by averaging over all fast angles. In this case, the only fast angle is $\lambda^\prime$ and the single averaging is given by
\begin{equation}\label{eq:averaging_disturbing_function}
	\epsilon \mathcal{H}_1 = \cfrac{1}{2\pi} \int_{0}^{2\pi} \mu \left(\cfrac{1}{\left| \bm{r}-\bm{r}^\prime \right|} - \cfrac{\bm{r}\cdot\bm{r}^\prime}{{\bm{r}^\prime}^3}\right) \mathrm{d} \lambda^\prime,
\end{equation}
where $\mu$ is the mass of the perturber, $\bm{r}$ the heliocentric position vector of the asteroid, $\bm{r}^\prime$ that of the perturber. It is worth noting that the averaged disturbing term~\eqref{eq:averaging_disturbing_function} has the same form and physical meaning with the ponderomotive potential \citep{Namouni:1999vp, 2013CeMDA.117..405M}.

In the following calculation, we always assume $\mu = 0.001$, which corresponds to the Sun-Jupiter three-body system.
\section{phase-space structure of retrograde 1:1 MMR}\label{sec:3}
Now we focus on the retrograde 1:1 resonance ($k=k^\prime=1$), which is the most special and complicated one among all the retrograde resonances. \citet{2013CeMDA.117..405M} has shown us that the order of the retrograde resonance is $\left|k+k^\prime\right|$ rather than $\left|k^\prime - k\right|$ in the prograde case. Therefore retrograde 1:1 resonance is of order 2, and it is the lowest possible order of the resonance for the retrograde orbits. Moreover, considering the geometry of a retrograde and a prograde orbit with the same semi-major axis, their orbits will inevitably intersect each other, no matter how small the eccentricity is.

\begin{figure}
	\includegraphics[width=0.9\columnwidth]{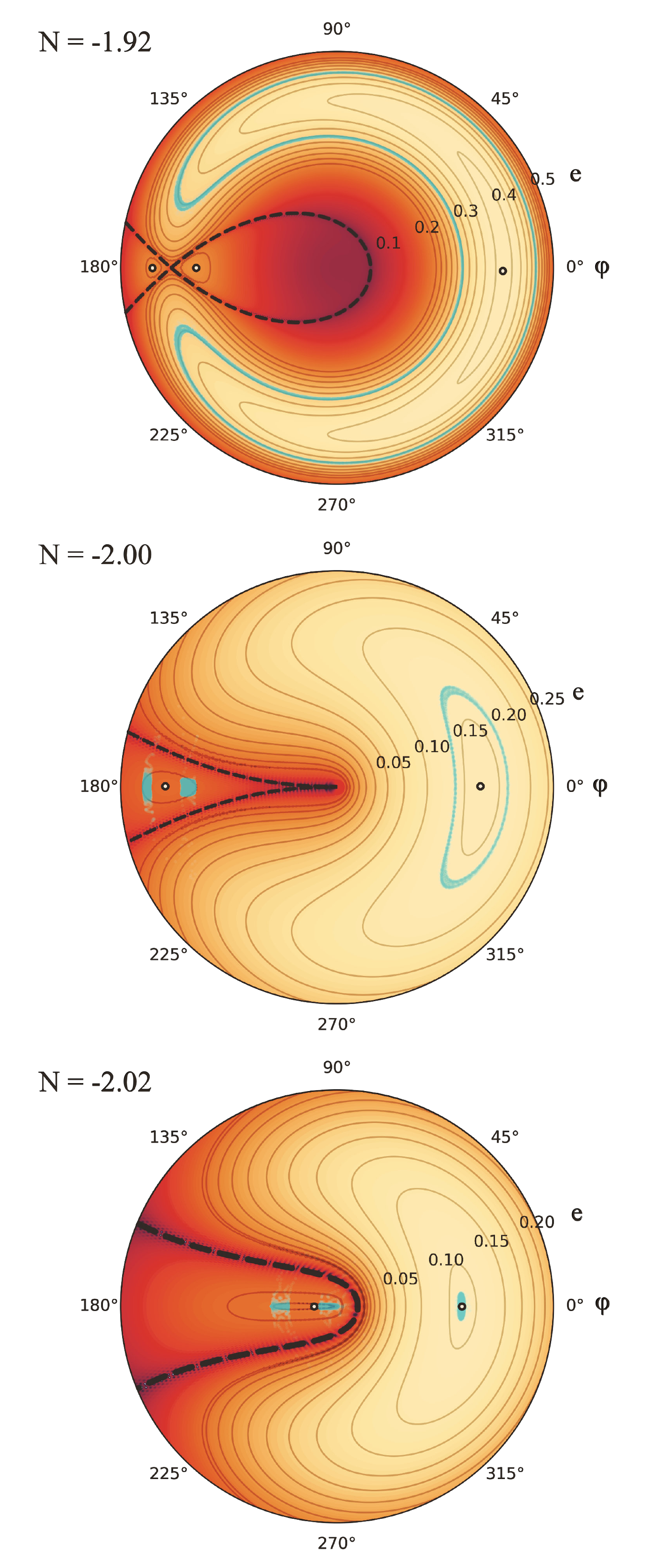}
	\caption{Retrograde phase portraits associated with the retrograde 1:1 mean motion resonance at different values of $N$, projected into $e-\varphi$ polar space. The variance of color and contours represent level curves of the $\mathcal{H}_{PC}$~\eqref{eq:planar_hamiltonian}. The thick black dashed lines denote collision curves with the perturber, while the hollow black dots mark the positions of the ideal equilibrium points of the resonance. The cyan curves in each panel represent the results of numerical integration in a circular restricted three-body model, whose variations of orbital elements and resonance angle are also plotted in Figure~\ref{fig:apocentric_curves} (apocentric libration).}\label{fig:phase_portraits_1_-1}
\end{figure}

In light of the analysis presented in the preceding subsection, we have the following critical angles for the 3D retrograde 1:1 mean motion resonance:
\begin{equation}\label{eq:1/-1_retrograde_angle_variables}
	\sigma = \cfrac{\lambda- \lambda^{\prime}-2\varpi}{2}, \quad \sigma_z = \cfrac{\lambda- \lambda^{\prime}+2\Omega}{2},
\end{equation}
The constant of motion is now
\begin{equation}\label{eq:1/-1_retrograde_constant}
	N =-2\Lambda + P +Q.
\end{equation}
If we restrict our motion to planar case, the second resonant angle $\sigma_z$ is neglectable and the constant $N$ can be simplified to $N = -\sqrt{a} \left(1+\sqrt{1-e^2}\right)$, which defines a curve in the $a-e$ plane. Next, we pick some different values of $N$ then do the averaging of~\eqref{eq:averaging_disturbing_function} and calculate the value of~\eqref{eq:planar_hamiltonian} on an $S-\sigma$ grid. Without loss of generality, we can map the level curves of $\mathcal{H}_{PC}$ from the $S-\sigma$ plane to another $e-\varphi$ plane, where $\varphi = (k+k^\prime) \sigma$, to get a better visual presentation. The corresponding $e-\varphi$ diagrams for three values of N ($-1.92, -2.00, -2.02$) are presented in Figure~\ref{fig:phase_portraits_1_-1}.

There is a lot of interesting information we can get from Figure~\ref{fig:phase_portraits_1_-1}. Above all, it is clear that the topology of $\mathcal{H}_{PC}$ for retrograde 1:1 resonance is somewhat similar to other mean motion resonances we have already known. More specifically, there are two groups of equilibrium points, one group at $\varphi = 0^\circ$ and another group at $\varphi = 180^\circ$, just like prograde mean motion resonances. Besides, the libration range around equilibrium point at $\varphi = 0^\circ$
is large while that around equilibrium point at $\varphi = 180^\circ$ is quite shallow.

The most distinctive feature of retrograde 1:1 resonance is that the collision curve always exists in the phase space, no matter the value of N. The collision curve slices the entire phase space into different regions. When $N\le-2$, the collision curve is not self-intersecting, thus there are two separated regions on the phase portrait (depicted in the second and third subfigures of Figure~\ref{fig:phase_portraits_1_-1}). We call the part on the right side of the collision curve the pericentric libration (libration around $\varphi = 0^\circ$), and left side the apocentric libration (libration around $\varphi = 180^\circ$). As mentioned before, two libration centers and the collision curve have also been discovered by \citet{2013CeMDA.117..405M} with ponderomotive potential. However, we notice that there is still one equilibrium point around $\varphi = 180^\circ$ and not mentioned in other studies. When $N>-2$, the phase space is sliced into three different regions (depicted in the first subfigure of Figure~\ref{fig:phase_portraits_1_-1}) and the number of equilibrium points at $\varphi = 180^\circ$ changes from one to two. The bifurcation of apocentric libration point has never been observed in any other resonances before, and it will be investigated in detail later.

\subsection{Pericentric libration}
Let us focus on the equilibrium point at $\varphi = 0^\circ$ first. It is clear that the results of numerical integration (cyan curves in Figure~\ref{fig:phase_portraits_1_-1}) around $\varphi = 0^\circ$ and the level curves of Hamiltonian correspond very well. Since the pericentric libration center is far from the collision curve, the particle is unlikely to get strong perturbation from the planet. So we can easily predict the dynamical evolution of pericentric libration with the phase portrait.



\begin{figure}
	\includegraphics[width=\columnwidth]{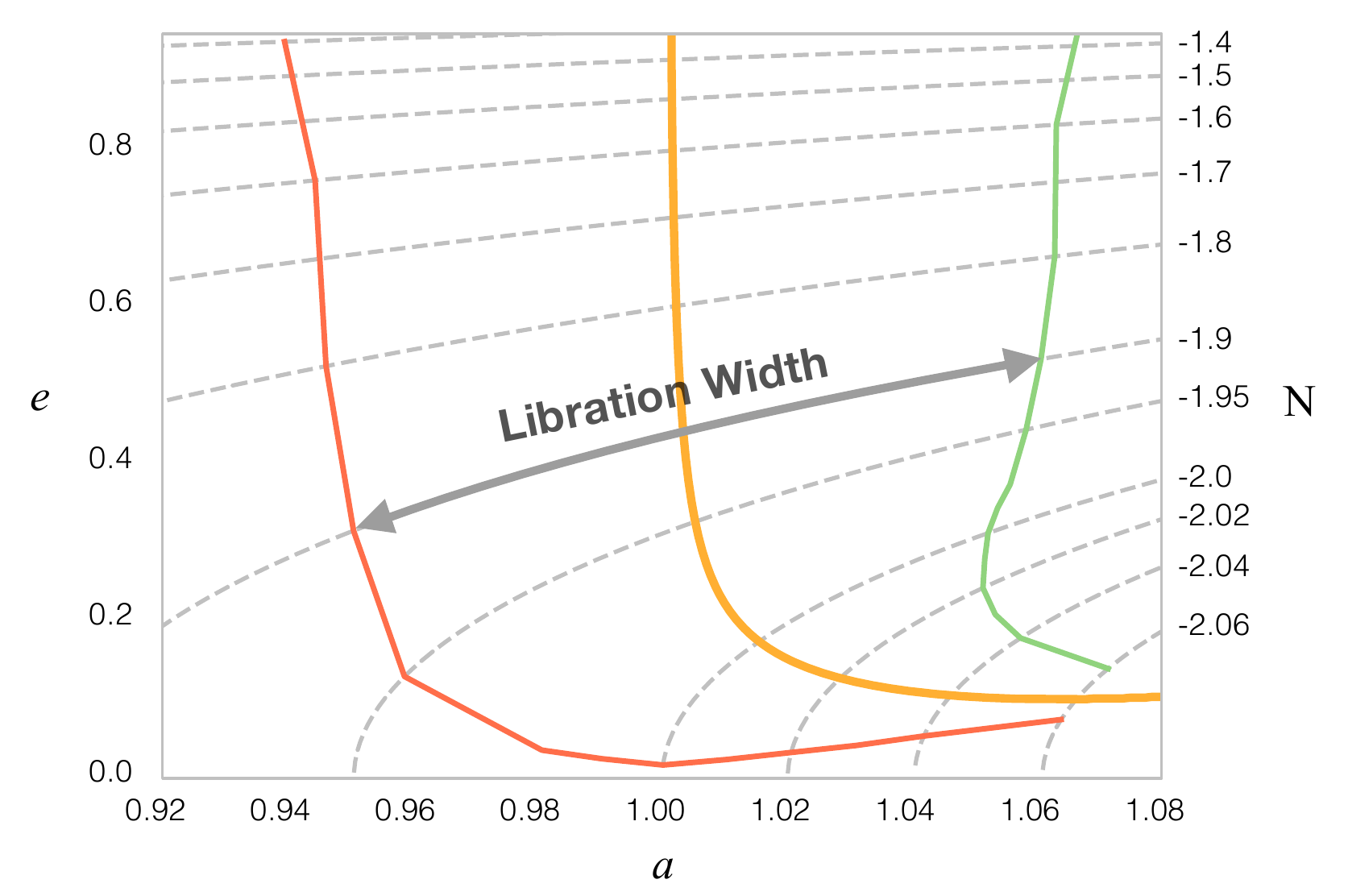}
	\caption{Exact resonance location around $\varphi = 0^\circ$ (solid orange line) and its maximum librational boundaries (red and green lines) on an $a-e$ plane. Some level curves of N (grey dashed line) and their corresponding values are shown on the plot. The libration width is the distance between two boundaries.}\label{fig:exact_resonance}
\end{figure}

The exact resonant location at $\varphi = 0^\circ$ on an $a-e$ plane is depicted in Figure~\ref{fig:exact_resonance}. Due to the existence of the collision curve, there is no well-defined separatrix on the phase portrait. Therefore the libration width, or the width of the resonance, is defined by the trajectory of largest librational amplitude that does not cross the collision curve \citep{Morbidelli:2002tn}. We evaluate the libration width of different values of $N$ and the results are shown in Figure~\ref{fig:exact_resonance}.

Unlike the case of prograde resonance, where the libration width decreases with increasing eccentricity above the planet-crossing curve \citep{2017AJ....154...20W}, we notice that the width of retrograde 1:1 resonance keeps growing with increasing eccentricity. However, our evaluation of libration width based on averaged perturbation is just a preliminary estimation. Direct numerical calculations are necessary to get accurate results.

\subsection{Apocentric libration}
It is easy to understand the dynamics around $\varphi = 0^\circ$. However, the dynamical evolution around $\varphi = 180^\circ$ is much more complex than that around $\varphi = 0^\circ$. Under the influence of the collision curve, the dynamical structure of apocentric libration has changed significantly. The real dynamics cannot be determined only by the level curves of averaged $\mathcal{H}_{PC}$.

As already mentioned above, when $N\le-2$, there is only one island of stable apocentric libration around $\varphi = 180^\circ$. Meanwhile, when $N>-2$, another new stable island appears on the phase portrait. Theoretically, one stable island represents one stable equilibrium point at its center. However, as the value of $N$ increases, the two stable islands become closer and closer to the collision curve. This means that apocentric libration will get strong perturbation from the close encounter with the planet, leading to the disruption of stable libration.

\begin{figure*}
	\includegraphics[width=\textwidth]{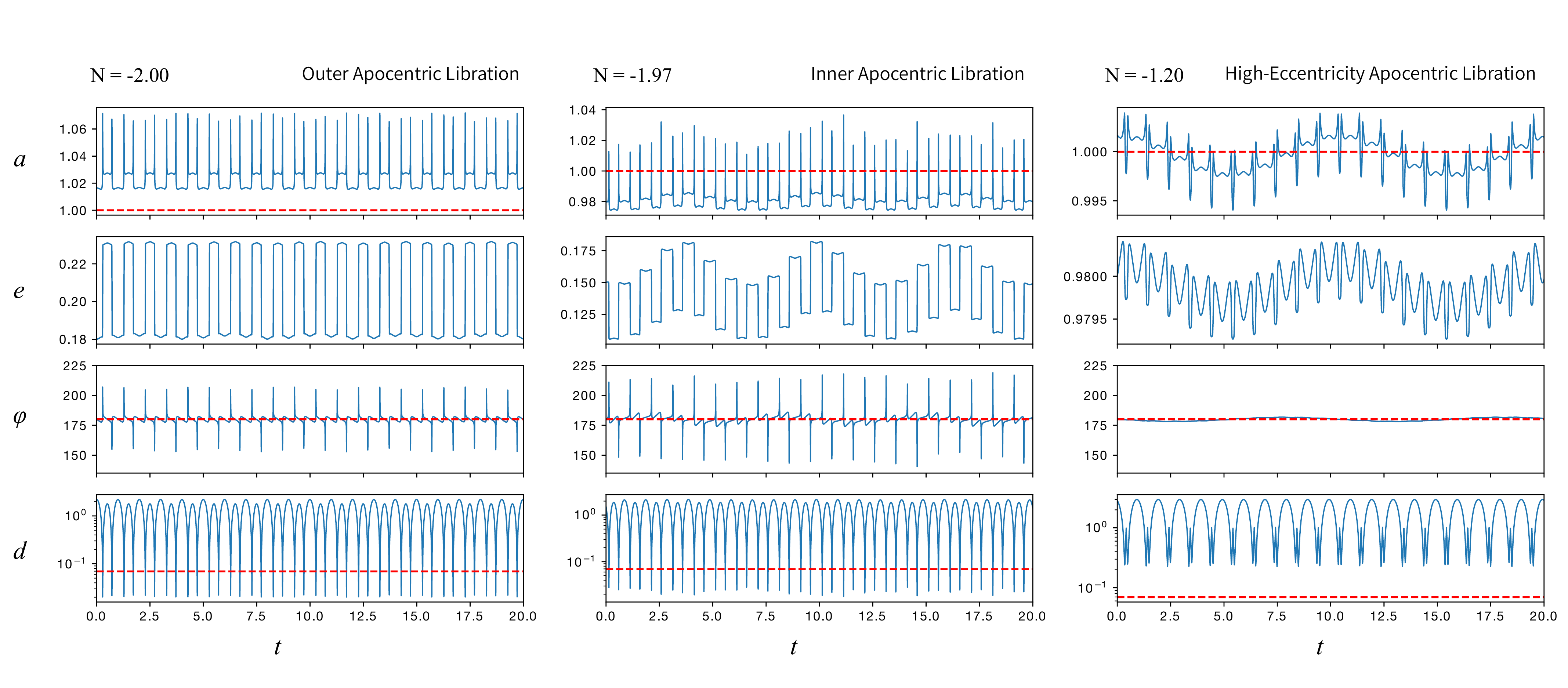}
	\caption{Dynamical evolution of semi-major axis $a$ (first row), eccentricity $e$ (second row), resonant angle $\varphi$ (third row) and distance from the perturber $d$ (fourth row) in the case of apocentric libration. The red dashed line in the first row represents $a=1$, in the third row represents $\varphi = 180^\circ$, while that in the fourth row represents $d = R_H$ (i.e. Hill Radius of the perturber). The left, middle and right panel correspond to the outer, inner and high-eccentricity apocentric libration defined in this paper, respectively. The values of N for numerical integration are shown on the top left of each panel. The significant jump of $a$, $e$, $\varphi$ and $d$ in outer and inner libration represents a close encounter with the planet.}\label{fig:apocentric_curves}
\end{figure*}

We do not observe any stable apocentric libration for $N$ between $-1.94$ and $-1.65$ in our numerical integration. For the other values of $N$, the stable island around $\varphi = 180^\circ$ is wide enough so that the libration becomes possible. To get a better understanding of this unique dynamical structure of retrograde 1:1 resonance around $\varphi = 180^\circ$, we classify the apocentric libration into the following three cases. For each case, we plot its $a$, $e$ and $\varphi$ evolution with time in Figure~\ref{fig:apocentric_curves}.

\begin{enumerate}
	\item Outer Apocentric Libration ($a$ librating above the exact resonance value $a_{res}$, plotted in the left panel of Figure~\ref{fig:apocentric_curves})
	\item Inner Apocentric Libration ($a$ librating below the exact resonance value $a_{res}$, plotted in the middle panel of Figure~\ref{fig:apocentric_curves})
	\item High-Eccentricity Apocentric Libration ($a$ librating around the exact resonance value $a_{res}$, plotted in the right panel of Figure~\ref{fig:apocentric_curves})
\end{enumerate}

It is helpful to recall the dynamical evolution of apocentric libration in the prograde case \citep[Chapter 8]{Murray:2000id}. The resonant angle librates around $180^\circ$ while the semi-major axis of the particle always librates above the exact resonance value $a_{res}$. Hence in our classification, all apocentric libration of prograde resonance should be outer apocentric libration. However, we found all three types of apocentric libration in the planar retrograde 1:1 resonance model.

Since the definition of $N$~\eqref{eq:1/-1_retrograde_constant} restricts that the value of $a$ must be larger than $1$ when $N\leq-2$, which means all the apocentric librations with $N\le-2$ are outer ones.  As for $-2\leq N \leq -1.94$, the outer and inner apocentric librations both exist, corresponding to the outer and inner islands around $\varphi = 180^\circ$ on the phase portrait respectively. As for $-1.94\leq N \leq -1.65$, both apocentric librations become highly unstable due to the strong and irregular perturbation from the planet. That is why they disappear in numerical integration.

\begin{figure}
	\includegraphics[width=\columnwidth]{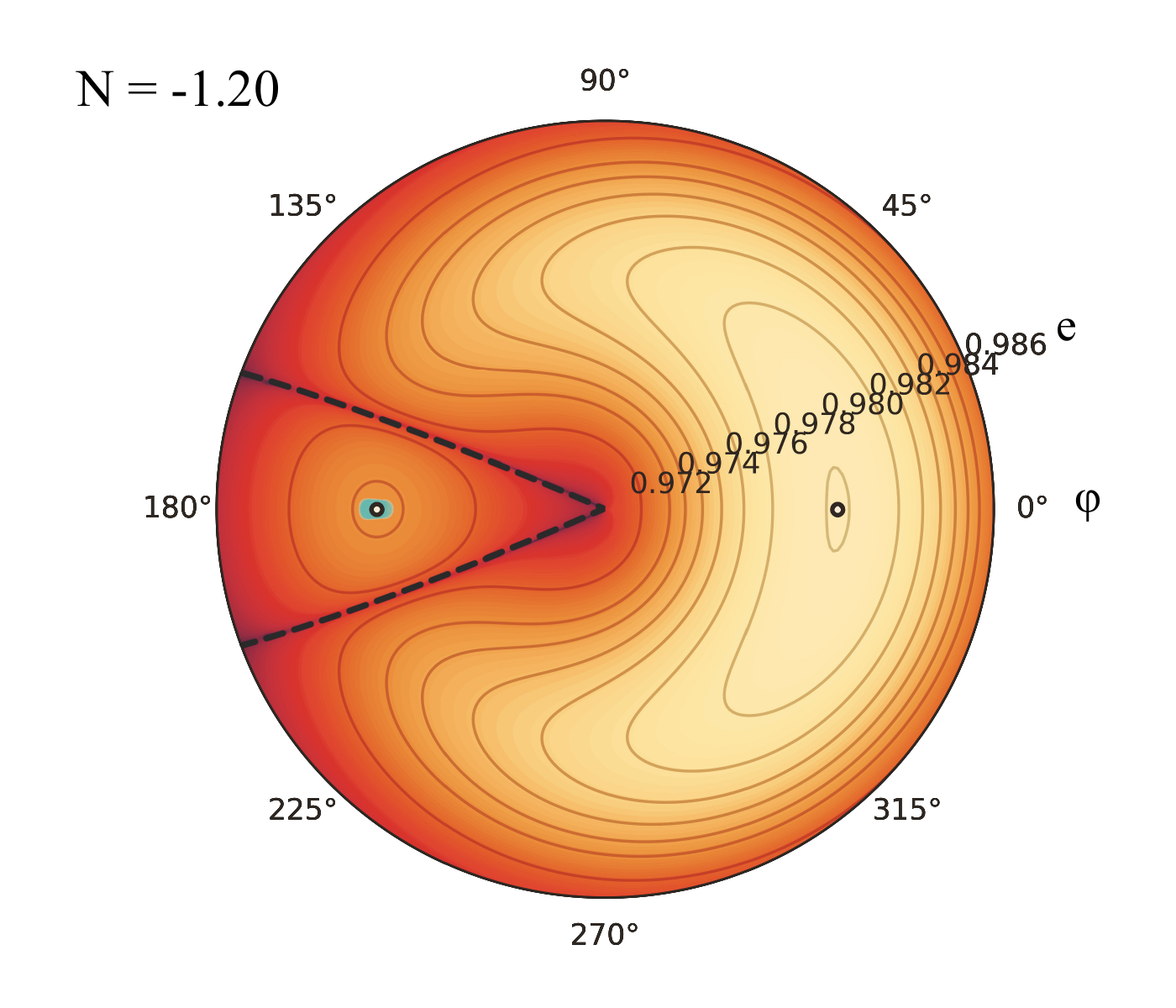}
	\caption{Phase portrait associated with the retrograde 1:1 mean motion resonance at $N=-1.2$. The basic meaning of this figure is the same as Figure~\ref{fig:phase_portraits_1_-1}, but we confine the plot range of eccentricity to $(0.971 , 0.986)$ for better visual presentation. The cyan curve around the libration center corresponds to the result of the right panel of Figure~\ref{fig:apocentric_curves}.}\label{fig:N=-120_portrait}
\end{figure}

While when $N \geq -1.65$ (or $e \geq 0.75$), the eccentricity of the retrograde orbit is high enough so that the phase portrait of retrograde 1:1 resonance changed again. One typical phase portrait of extremely high eccentricity is depicted in Figure~\ref{fig:N=-120_portrait}.

As shown in Figure~\ref{fig:N=-120_portrait}, the stable apocentric island regains its breadth, so that the influence of the collision curve become much weaker and the apocentric libration is stable again. In addition to the topological variation of phase portrait, it can be seen in Figure~\ref{fig:apocentric_curves} that the dynamical evolutions of outer and inner apocentric librations in retrograde 1:1 resonance are significant compared to the high-eccentricity apocentric libration. As shown in the left and middle panels of Figure~\ref{fig:apocentric_curves}, although the average of resonant angle $\varphi$ is very close to $180^\circ$, its peak value can jump above $200^\circ$. It is noteworthy that these jumps cannot be reproduced by our semi-analytical model since the fast angles have been removed through averaging. Nevertheless, the phase-space portrait indeed indicates the location of apocentric centers so it still supplies a good reference to the problem.

\begin{figure}
	\includegraphics[width=\columnwidth]{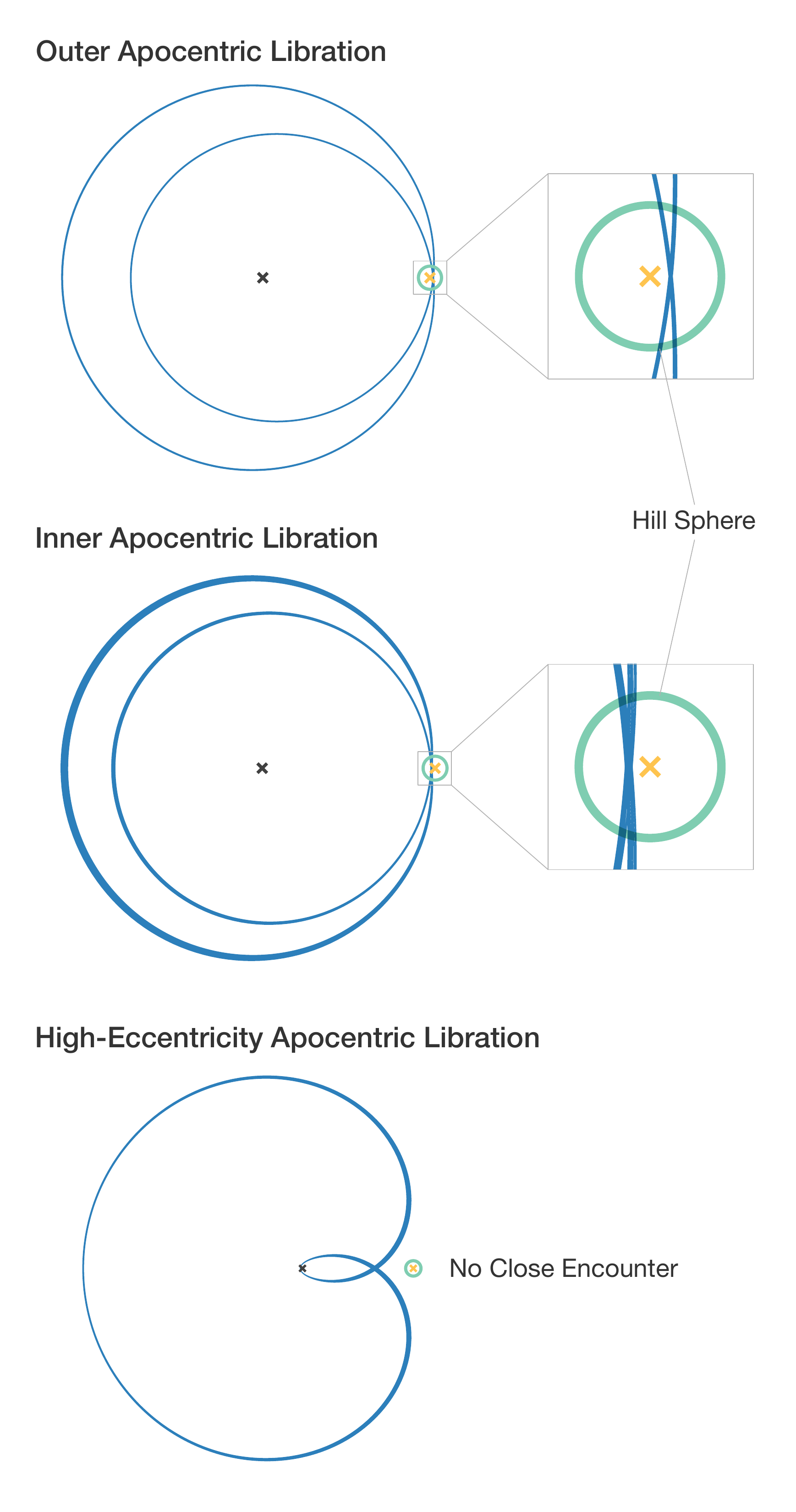}
	\caption{Orbits in the retrograde 1:1 resonance seen in the synodic frame. The three panels correspond to outer, inner and high-eccentricity apocentric libration respectively. The position of Sun is marked by black cross while the planet is marked by the orange cross. The green circle represents Hill sphere of the planet, which is used to evaluate the closest distance between the asteroid and the planet. The variations of orbital elements and resonant angle of each orbit are also depicted in Figure~\ref{fig:apocentric_curves}. Similar orbits in the synodic frame are also shown in \citet[Figure 4]{2013CeMDA.117..405M}.}\label{fig:apocentric_detail}
\end{figure}

The significant jumps seen on the semi-major axis are caused by the close encounters with the planet. This is evidenced by the trajectories of test particle in the synodic frame for three types of apocentric librations in Figure~\ref{fig:apocentric_detail} where a clear correlation of the orbital elements jumps and the relative distance to the perturber is seen.

As can be clearly seen in Figure~\ref{fig:apocentric_detail}, in the case of outer and inner apocentric libration, the trajectory of the test particle indeed enters the Hill sphere from two slightly different directions. One direction boosts the semi-major axis of the orbit while another one decreases that (also shown in Figure~\ref{fig:apocentric_curves}). In one synodic period, the semi-major axis of the orbit gets raised once and also gets reduced once, the absolute value of two changes are nearly equal, so the orbit is still in resonance.


For high-eccentricity apocentric libration, the trajectory of the test particle is always outside the Hill sphere of the planet. Without close encounter, the particle can not get strong perturbation from the planet, hence the variations of $a$, $e$ and $\varphi$ are all subtle. As shown in Figure~\ref{fig:N=-120_portrait} with the cyan curve, the apocentric libration with high eccentricity can be very close to the theoretical equilibrium point. This is, however, not the case for outer and inner libration, whose numerical dots are scattered on both sides of the equilibrium point in Figure~\ref{fig:phase_portraits_1_-1}, no matter how close we put the initial conditions to the equilibrium point.

\section*{conclusions}
In this paper, we constructed a one-degree integrable approximation model for any retrograde mean motion resonances in a planar circular restricted three-body problem, following a similar procedure proposed by \citet{Morbidelli:2002tn}. When dealing with the disturbing term, we abandoned the classical series expansion method and carried out the averaging process numerically over synodic angle $\lambda^{\prime}$ in closed form.

Then we focused on a specific retrograde mean motion resonance, the retrograde 1:1 resonance, and elucidated its phase-space structure on an $e-\varphi$ polar plane. We noticed that its phase portrait varies significantly when the constant of motion $N$ changes, leading to the bifurcation of equilibrium point around $\varphi = 180^\circ$. Due to the influence of the collision curve, we found that stable apocentric libration only survives when the eccentricity is small or extremely high (more specifically when $-1.94 < N < -1.65$). We analyzed three types of apocentric librations, the inner one, the outer one and the high-eccentricity one, and found that the significant jumps in orbital elements are caused by close encounters with the perturber.


\acknowledgments
This work is supported by the National Natural Science Foundation of China (Grant No.11772167).

\bibliography{ref}
\bibliographystyle{aasjournal}

\end{document}